\documentstyle[11pt,newpasp,twoside,epsf]{article}

\reversemarginpar

\begin{document}
\title{A Fast Search Technique for Binary Pulsars}
\author{Scott M.~Ransom}
\affil{Harvard-Smithsonian Center for Astrophysics, 
  60 Garden St., MS 10, Cambridge, MA 02138, USA}

\begin{abstract}
  I describe a computationally simple, efficient, and sensitive method
  to search long observations for pulsars in binary systems.  The
  technique looks for orbitally induced sidebands in the power
  spectrum around a nominal spin frequency, enabling it to detect
  pulsars in high- or low-mass binaries with short orbital periods
  ($P_{orb} \la 5 \mathrm{\; h}$).
\end{abstract}

\noindent
Pulsars in binary systems experience orbital accelerations which
Doppler shift the pulsar spin frequencies.  This Doppler
smearing, and the huge parameter space we must search because of it,
makes the discovery of all but the longest-period binaries virtually
impossible (e.g., Johnston \& Kulkarni 1991).
  
The new technique, which I call {\em phase modulation searching},
greatly reduces the size of the search parameter space, with only a
modest decrease in sensitivity compared to a fully coherent search.
These searches rely on the fact that if the observation time is longer
than the orbital period ($T_{obs} > P_{orb}$) the orbital Doppler
shifts effectively phase modulate the pulsar spin frequency.

Phase modulation results in a family of sidebands, evenly spaced by
the orbital period, around the intrinsic pulsar signal in the
frequency domain.  These sidebands are effectively short periodic
sequences in the full power spectrum and are detectable using short
FFTs or sums of powers.  When $T_{obs} > P_{orb}$ these periodic
sidebands occur for all combinations of Keplerian orbital elements,
which reduces the binary portion of the search parameter space to only
2 parameters:  the modulation period, $P_{orb}$, and amplitude,
$\Phi_{orb} = 2 \pi x f_{psr}$ ($x$ is the projected semi-major
axis of the orbit and $f_{psr}$ is the pulsar spin frequency).

The simplicity and efficiency of phase modulation searching allows the
analysis of very long time-series (e.g., satellite observations of a
source over many days or year-long data sets from gravitational wave
detectors) which greatly improves the chances of detecting weaker
sources.  Data analysis using these and other advanced Fourier
techniques is ongoing on several long radio observations of globular
clusters (see Figure 1) and on various RXTE observations.

\acknowledgments I would like to thank both Steve Eikenberry and Jim Cordes
for useful ideas and discussions related to this work.

\begin{figure}[ht]
\setlength{\unitlength}{1in}
\begin{picture}(0.0,4.1)
\put(-0.2,4.3){\includegraphics{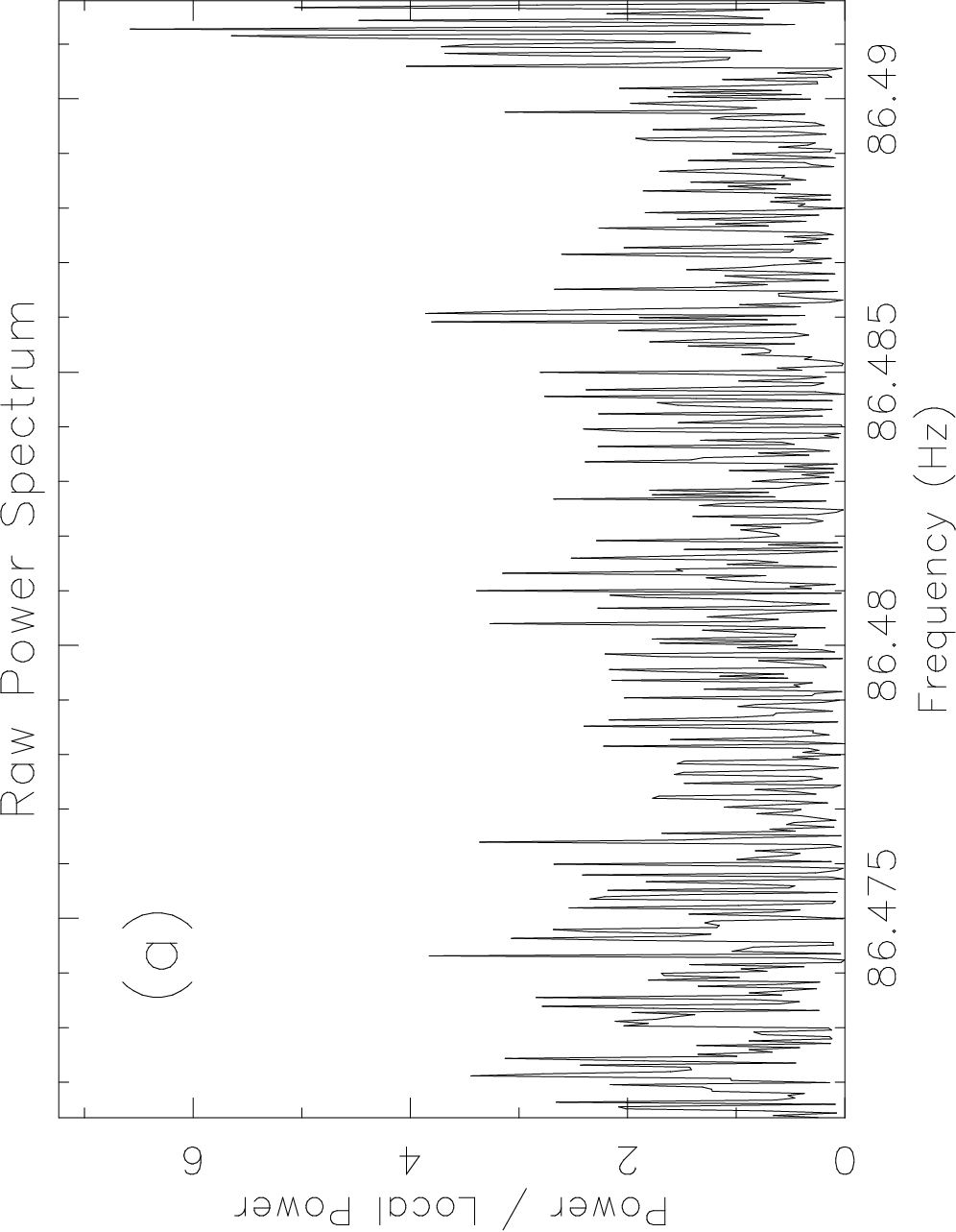}}
\put(2.5,4.3){\includegraphics{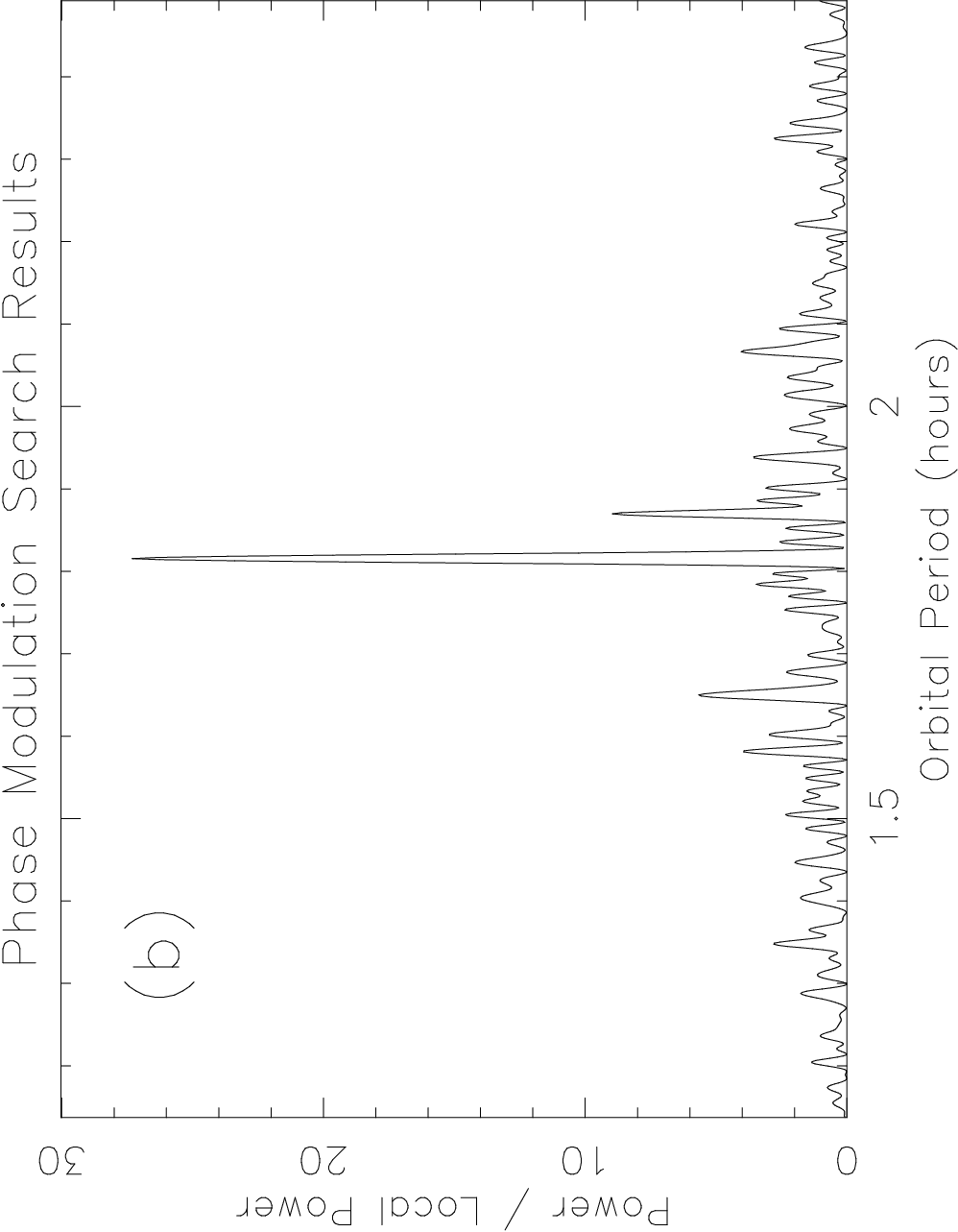}}
\put(-0.2,2.2){\includegraphics{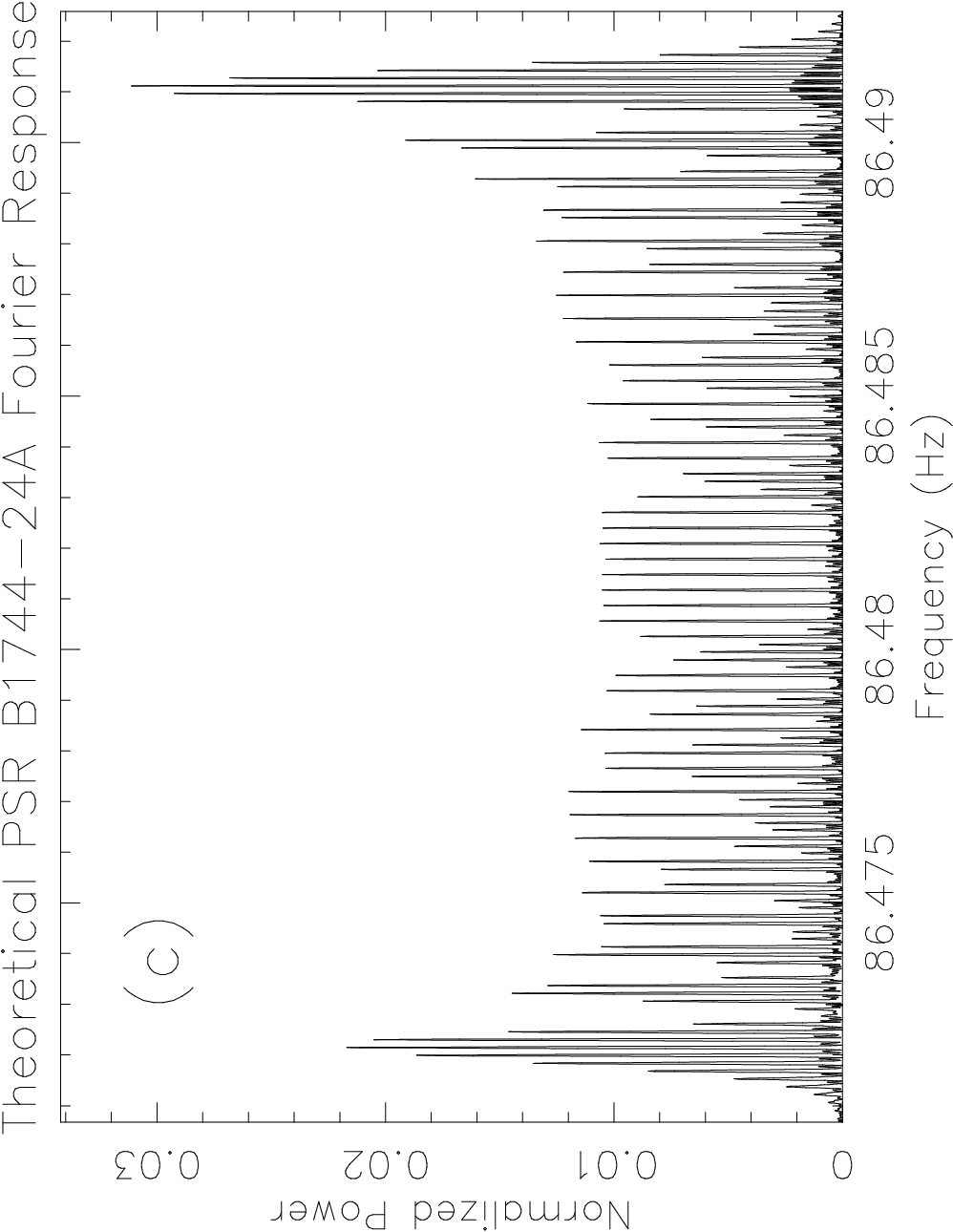}}
\put(2.5,2.2){\includegraphics{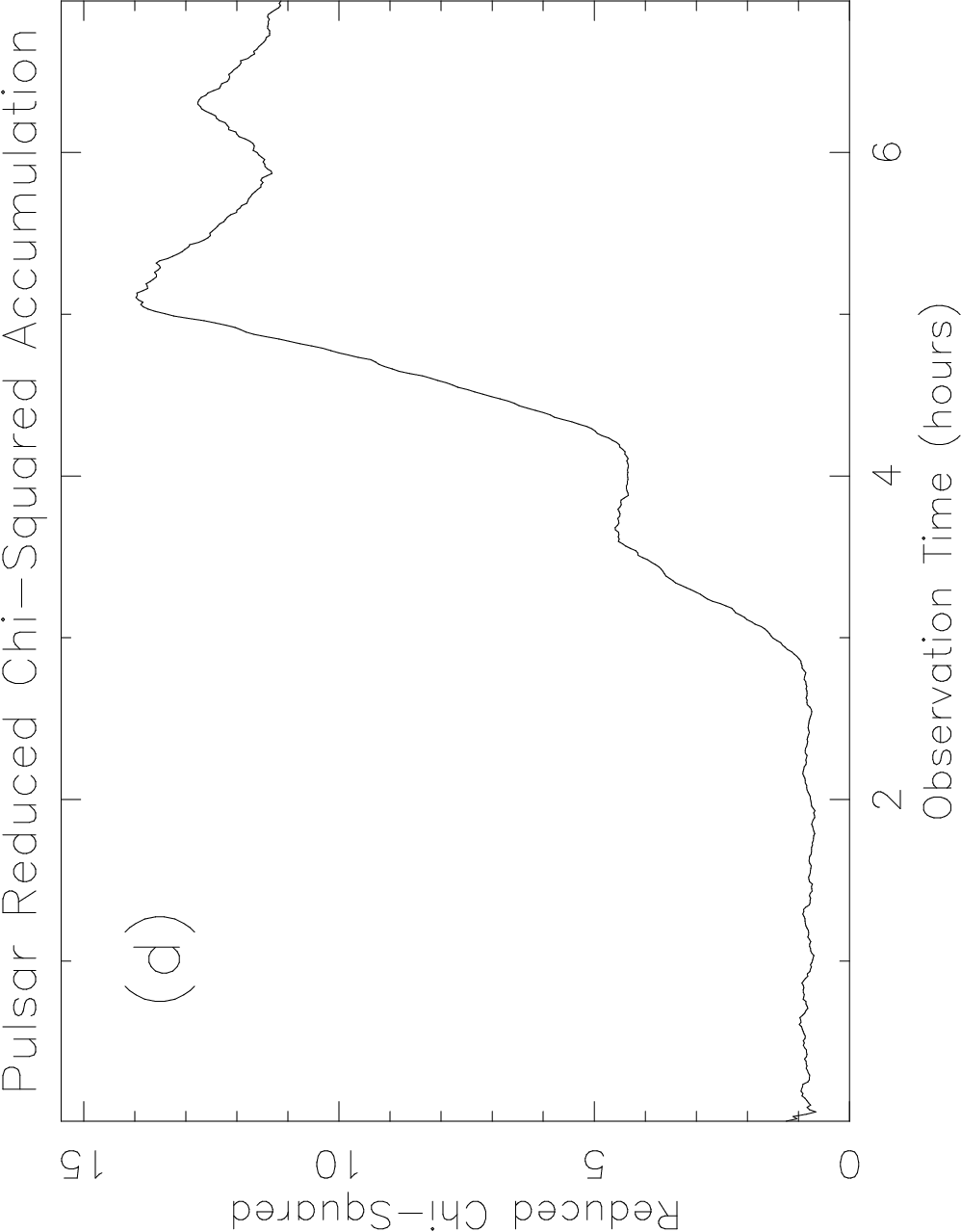}}
\end{picture}
\caption{
  Results from a phase modulation search of a 7 hour observation of
  the globular cluster Terzan5 taken with the Parkes Multibeam system.
  (a) shows a section of the power spectrum from the full observation
  centered on the spin frequency of the eclipsing and ``disappearing''
  binary millisecond pulsar PSR B1744-24A $(P_{orb} = 1.82 \mathrm{\;
    h})$ (see Nice \& Thorsett 1992).  All power levels are consistent
  with being due to noise.  (b) shows the $\sim7\,\sigma$ detection of
  the pulsar by taking a short FFT of (a).  Note that we detect the
  pulsar at the accurately determined {\em orbital} period instead of
  the spin period.  The detection results from the fact that buried
  beneath the noise in (a) are the periodic sidebands caused by the
  orbital phase modulation of the pulsar spin frequency.  (c) shows
  what these sidebands would look like for PSR B1744-24A with no noise
  if it were present during the full observation.  The total number of
  sidelobes is approximately equal to $2\Phi_{orb} \sim130$.  (d)
  shows the accumulation of reduced $\chi^{2}$ while folding the
  pulsar signal at the known ephemeris.  The pulsar is ``on'' for only
  $\sim2$ non-continuous hours during the observation.  The fact that
  we detect the pulsar using a phase modulation search, even though
  the orbit only ``modulates'' the spin frequency for approximately
  one non-continuous orbital period, demonstrates the robustness of
  the search technique.  If the pulsar had been present during the
  full observation, the sidelobes would have been sharper and
  stronger, resulting in a much more significant detection.}
\end{figure}

\end{document}